# Fabrication of $Ge_2Sb_2Te_5$ metasurfaces by direct laser writing technique

D.V. Bochek[1], K.B. Samusev[1,2], D.A. Yavsin[2], M. V. Zhukov[1,3], M.F. Limonov[1,2], M.V. Rybin[1,2], I.I. Shishkin[1], A.D. Sinelnik[1]

[1] ITMO University, Faculty of Physics and Technology, St. Petersburg 197101, Russia

[2] Ioffe Institute, St Petersburg 194021, Russia

[3] Institute for Analytical Instrumentation RAS, Laboratory of scanning probe microscopy and spectroscopy, St. Petersburg 198095, Russia

## Abstract

We experimentally demonstrate fabrication of tunable high contrast periodic fishnet metasurfaces with 3 μm period on 200 nm thick $Ge_2Sb_2Te_5$ films sputted onto glass and sapphire substrates using direct laser writing technique. We find that the use of sapphire substrate provides better accuracy of metasurface segments due to high thermal conductivity. The advantages of the demonstrated method consist in its simplicity, rapidity, robustness, and the ability of tuning of fabricated structures. This is of crucial importance for the creation of robust and tunable metasurfaces for applications in the field of telecommunications and information processing.

## INTRODUCTION

Metasurfaces have attracted great interest in recent years since they provide versatile possibilities for the manipulation of optical waves, including amplitude, phase, polarization, and angular momentum promising an effective method for miniaturization and integration of optical devices [1-3]. At the same time, reconfigurable metamaterials have attracted considerable interest for their ability to offer controllable and adaptive optical properties 'on demand' [4, 5]. Phase-change media offer the possibility to realize reconfigurable all-dielectric metamaterials and metasurfaces because their phase state can be reversibly switched through thermal cycling in non-volatile way, leading to large changes in optical and electrical properties.

$Ge_2Sb_2Te_5$ (GST) is the most commercialized phase-change material, and it is employed as the active material in information storage devices such as rewritable optical disks and memory devices [6]. GST can switch between an amorphous and a crystalline states and show significant change in refractive index from a value of $n_a = 3.9 + 4.2i$ in the amorphous phase to $n_c = 4.3 + 2.0i$ in the crystalline phase at a wavelength of 0.73 μm [7]. GST has a crystallization temperature of around 150 °C and melting point about 700 °C. Using these properties of GST in combination with resonance properties, which can be determined by the structure of metasurfaces, one makes it possible to make a breakthrough in controlling light fluxes at the micro level.

The reverse crystalline-amorphous phase transition, as well as the direct one, is realized due to heating of GST film up to 700°C. Thereby the high thermal conductivity of the substrate is a significant factor for the existence of an inverse phase transition. For example, at room temperature sapphire has a thermal conductivity perpendicular to the C axis of the crystal lattice κ = 25.2 W/(m*K) in comparison with glass where κ = 0.8 W/(m*K). For fast cooling of the GST film, the cooling rate must be at least $10^{11}$ K/s, the calculation shows that the thermal conductivity of sapphire will achieve the required cooling rate for the reverse phase transition. Moreover, optical loss in sapphire is approximately the same as in glass in the telecom range and substantially smaller in the wavelength range of 2-4 μm, which is of great importance for various applications in medicine and biology. The silicon substrate has much greater thermal conductivity κ = 149 W/(m*K) and is commonly used in non-transmissive metasurfaces [8].

Metasurfaces structured from GST films can be fabricated using several nanofabrication technologies [9], such as nanoimprint lithography [10], electron beam lithography [11], ion beam etching [12] and direct laser writing (DLW) technique [13, 14]. At present, one of the most promising additive technology areas for creating various structures and entire devices in the sub-wavelength range is fs-laser lithography. DLW technique through Ti:Sa femtosecond pulses at around 780 nm wavelength causes temperature phase transitions of GST or its ablation, and the creation of 2D microstructures of arbitrary shape with a resolution of up to 200 nm. The advantage of the DLW technique over ion beam etching and electron beam lithography, lies in its simplicity and manufacture speed. Single-pass laser drawing under velocity of hundreds of micrometers per second provides the creation of entire metasurfaces with high contrast elements of subwavelength dimension in a few minutes [15].

To date, the optical properties of GST metasurfaces are widely studied [16-19]. The optical induced phase transition of GST has been proposed and used in many publications for manufacturing of variety of optical metadevices. Firstly, design a near-infrared tunable transmittive metalens utilizing GST phase change materials as the meta-atoms was presented in the work [20]. The focusing efficiency of the metalens in amorphous state at wavelength, 1.55 μm, was 16 times of the crystalline one. Reconfigurable bichromatic and multi-focus Fresnel zone plate were experimentally demonstrated in the work [21]. Such component was written, erased and rewritten as two-dimensional binary or greyscale patterns into a nanoscale film of GST material by inducing a refractive-index-changing phase transition with trains of femtosecond pulses. In paper [22] a rectangular antenna-based metasurface based on the phase-changing material GST with high transmittance and polarization conversion rate was proposed. Before the GST phase transition, the metasurface operates as a quarter-wave plate in the wavelength range of 10.0–11.9 μm. After the GST turning into crystalline state, the metasurface operates as a half-wave plate in the wavelength range of 10.3–10.9 μm. A reconfigurable metasurface made of GST was experimentally demonstrated in the 1.55 μm range and has been optimized to switch from highly transmissive (80%) to highly absorptive (76%) modes with a 7:1 contrast, when thermally transformed from the amorphous to crystalline state [23].

In this study, we present metasurfaces fabricated by laser scrabbing of GST in the form of periodic arrays of square islets of ~2.5 μm$^2$ in size with 3 μm period on 200 nm thick GST films thick on two types of substrates: from glass and from sapphire, using the DLW technique. Further, the samples were experimentally investigated by atomic force microscopy (AFM) and scanning electron microscopy (SEM). As a result of the comparative analysis, we conclude that proposed method guarantees quick manufacturing of high quality metasurface and of the use of sapphire substrate provides better accuracy of metasurface segments.

## Fabrication method of GST metasurface

To fabricate GST films onto silica or sapphire substrates we applied the laser electrodispersion technique. The technique consists of splashing of submicrometer droplets from the target surface by an intense laser pulse and their subsequent cascade fission. The targets were made of polycrystalline GST, synthesized from special-purity elements Ge, Sb, and Te. We used the Nd:YAG laser with wavelength of 1064 nm, a pulse repetition rate of 30 Hz, a focused spot diameter of 1 mm and power density of 109 W/cm$^2$ which was enough for an optical breakdown of the vapor to give a laser torch and evaporate the target material. In the vicinity of laser torch submicrometer droplets are divided into nanometer size ones under electrostatic forces and then deposited onto a substrate in the amorphous state. Transmission electron microscope (TEM) shows that the deposited films are composed of GST grains with sizes of 2-5 nm.

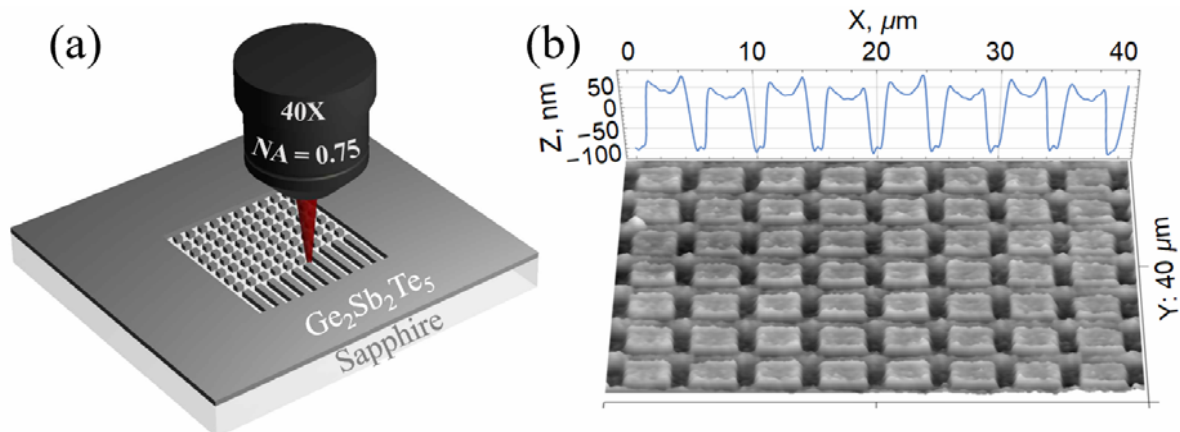

Fig. 1. (a) Schematic illustration of the laser scrabbing process of the 200 nm GST film through direct laser writing technique, (b) AFM image of the fabricated fishnet metasurface with 5 μm period, Fluence W = 0.047 J/cm$^2$ and 8×10$^5$ pulses per μm. Blue line on a graph above shows the cross-section averaged profile of all squares.

We used DLW technique to fabricate metasurfaces from an GST film on a glass and sapphire substrate. The schematic illustration of the laser scrabbing process is illustrated in Fig. 1(a). The 50-fs pulses centered at around 780 nm wavelength at a repetition frequency of 80 MHz are derived from a TiF-100F laser. Laser radiation was focused on the film with a 40X microscope objective with numerical aperture NA = 0.75. The sample was mounted on a horizontal two-coordinate motorized linear air-bearing. The objective was fixed on third vertical axis of translator that is necessary for focus setting. The motion of all three air translators was controlled by a computer with nanoFab GmbH. Fig. 1(b) demonstrates an

AFM image of the fabricated fishnet metasurface and the blue curve on the graph in the graph is the cross-section averaged over all squares.

## Comparison of metasurfaces on sapphire and glass substrates

In our first experiment we focused on the selection of optimal parameters for fabrication of GST fishnet metasurfaces on glass and sapphire substrates using the laser scrabbing method. For this purpose, it is necessary to find out the conditions under which the film is pierced to the bottom, since this will provide a thinner cutting line and increase the accuracy of the metasurface segments. Thus, we made test arrays of holes on the films, gradually changing the fs-laser power and duration of exposure over the two axes. Fig. 2. shows SEM images of obtained arrays in the GST films deposited on glass (Fig. 2(a)) and sapphire (Fig. 2(b)) with fluence range from 0.042 J/cm$^2$ to 0.06 J/cm$^2$, and exposure range from 60 to 300 pulses. Fig. 2 demonstrates that the sets of holes form practically triangular matrices in the "time-power" coordinates; therefore, the fabrication of fishnet structures is reasonable with parameters lying near the matrix diagonal. The main conclusion, which is especially well shown in Figures 2(c) and 2(d), is that the sapphire substrate ensures the formation of a smooth shape holes not strongly dependent on changes in the fabrication parameters.

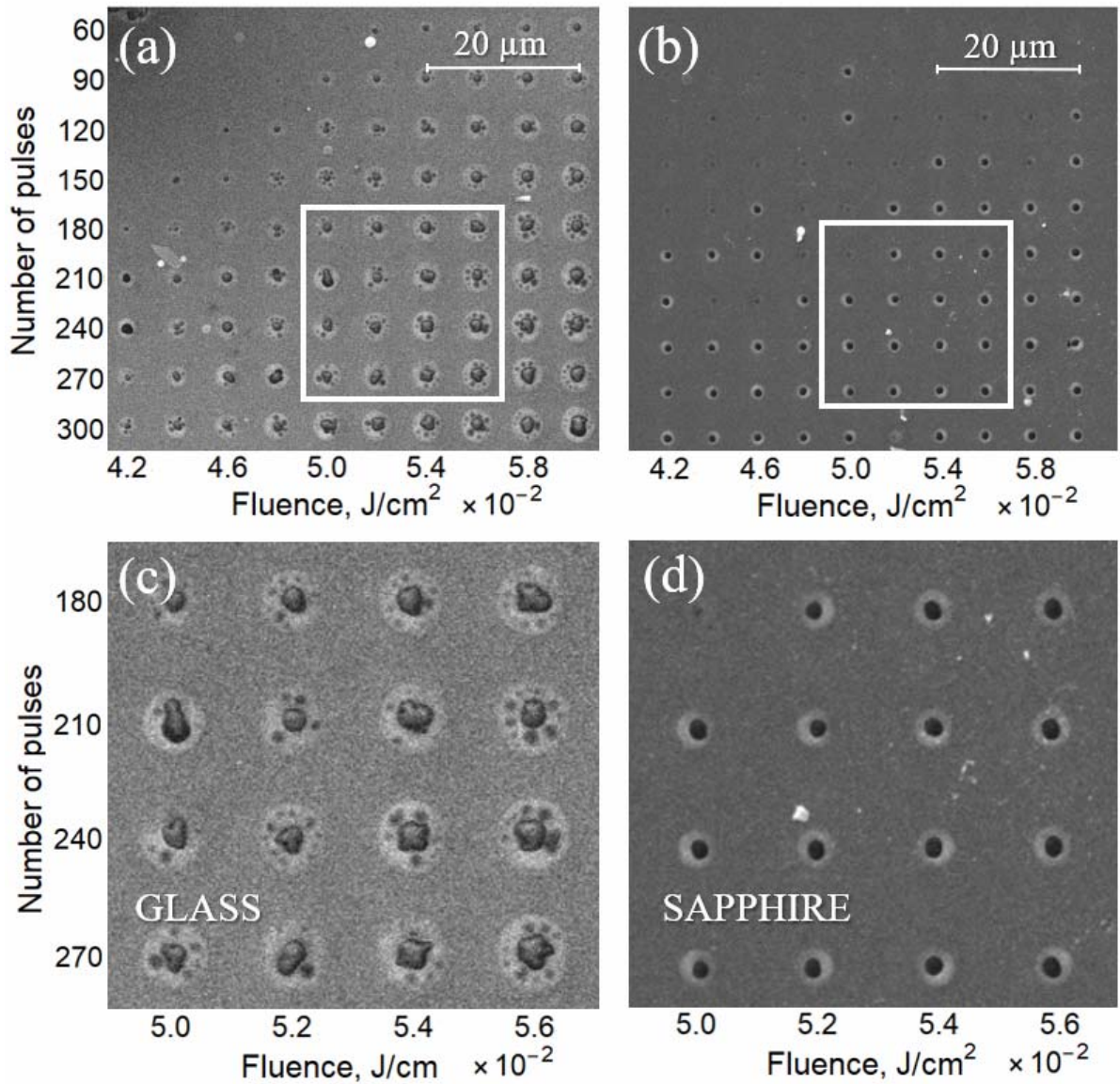

Fig. 2. SEM image of holes in GST film on the (a) – glass and (b) – sapphire substrates fabricated with fluence range of 0.042 J/cm$^2$ – 0.06 J/cm$^2$, and exposure range from 60 to 300 pulses. (c) – Magnified region of the SEM image outlined in (a), and (d) – magnified region of the SEM image outlined in (b).

For a more rigorous analysis, the surface topography of the obtained holes (Fig. 3(a, b)) was studied using an atomic force microscope (AFM). AFM is a powerful tool providing high resolution quantitative information about the surface. In Fig. 3(c, d) we demonstrate cross-section profiles of the ablated holes on a glass and sapphire substrates, fabricated with fluence range of 0.056 - 0.062 J/cm$^2$ and exposure of 270 – 300 pulses. It is clearly seen that holes on the glass substrate have irregular shape and depth, whereas holes on sapphire have smooth shape which does not strongly depend on slight deviations of fluence and exposure.

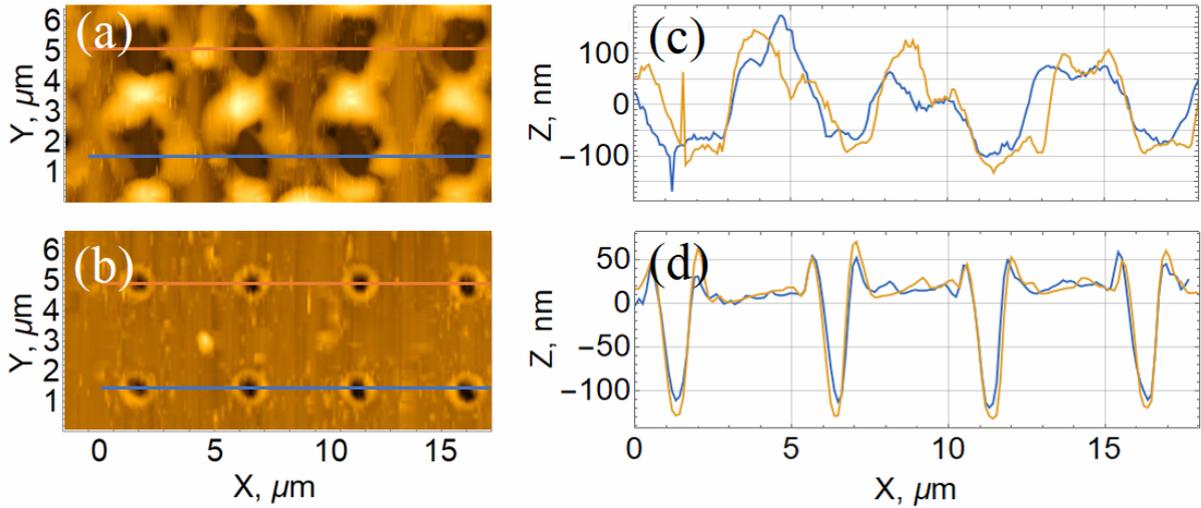

Fig. 3. (a) and (b) - AFM images of the ablated holes on a glass and sapphire substrates, respectively, fabricated with fluence range of 0.056 - 0.062 J/cm$^2$ and 270 – 300 pulses. Plots (c) and (d) show two cross-section profiles each, along the lines of the corresponding color drawn in (a) and (b).

Further, we created fishnet metasurfaces by direct fs-laser scrabbing of the GST film on both glass and sapphire substrates. Fig. 4 shows SEM images of the 3 μm period fishnet structures fabricated in the most optimal modes for each substrate, which ensure complete removal of material and a clear shape of the metasurface segments as a result of the scrabbing process. For glass substrate (Fig. 4(a)) the translator speed was 100 μm/s ($8\times10^5$ pulses per μm), the laser power was 90 mW (fluence is $4.2\times10^2$ J/cm$^2$) and for sapphire substrate (Fig. 4(b)) the translator speed was 100 μm/s, the laser power was 95 mW ($4.4\times10^2$ J/cm$^2$). The time to create one well-ordered metasurface of about 20x20 μm$^2$ in size and the number of square elements of the order of 50 was approximately 10 seconds. It is clearly seen that the use of sapphire substrate provides substantially more precise square shape, which is a consequence of the rapid heat dissipation in sapphire.

From our experiments, it can be unambiguously concluded that the use of a sapphire substrate has many advantages such as a thinner cutting line, as well as its clearer contour. Combined with the fact that sapphire is transparent for a wider wavelength range (0.17-5.5 μm), it becomes obvious that this material is better suited for creating GST structures by laser scrabbing.

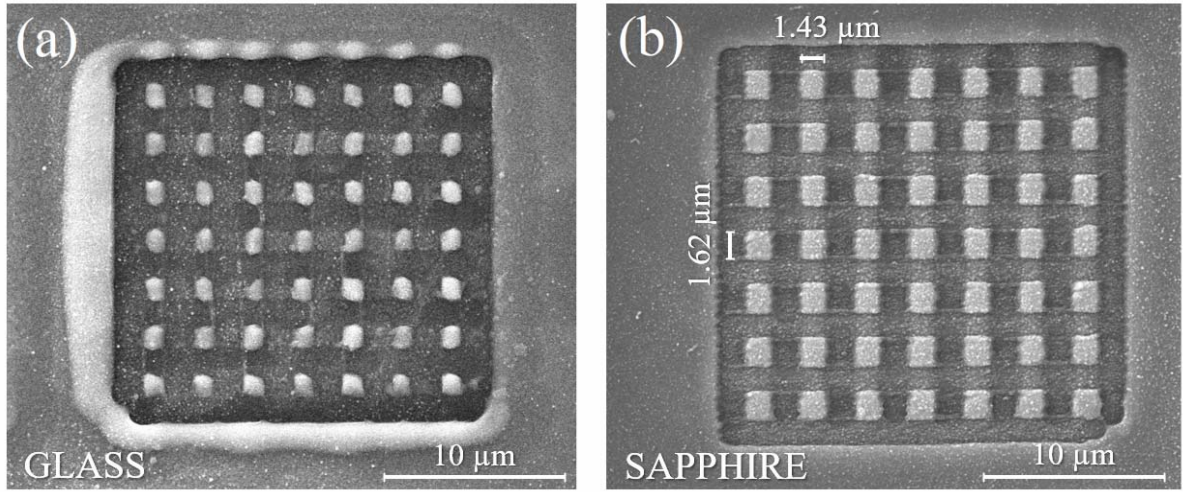

Fig. 4. SEM images of the 3 μm period fishnet structures on glass (a) and sapphire (b) substrates under the same fabrication parameters.

## Different fabrication regimes

It is possible to create the metasurfaces with various segment sizes at the same period by changing power of fs-laser or number of pulses per μm. Fig. 5 (a-l). demonstrates SEM images of 20x20 μm fishnet structures with 3 μm period fabricated with gradually changing fs-laser power and velocity of linear translators. The fluence value was varied in the range of 0.04 - 0.07 J/cm$^2$ and the translation velocity range was of 50-150 μm/s. It is seen in magnified region of the SEM image outlined in Fig. 5(e), that at the lowest fluence value the cutting line is thin enough for metasurface segments sticking together (Fig. 5(m)) and, conversely, with a fluence value from 0.07 J/cm$^2$ the squares become too small and indecipherable (Fig. 5(p)).

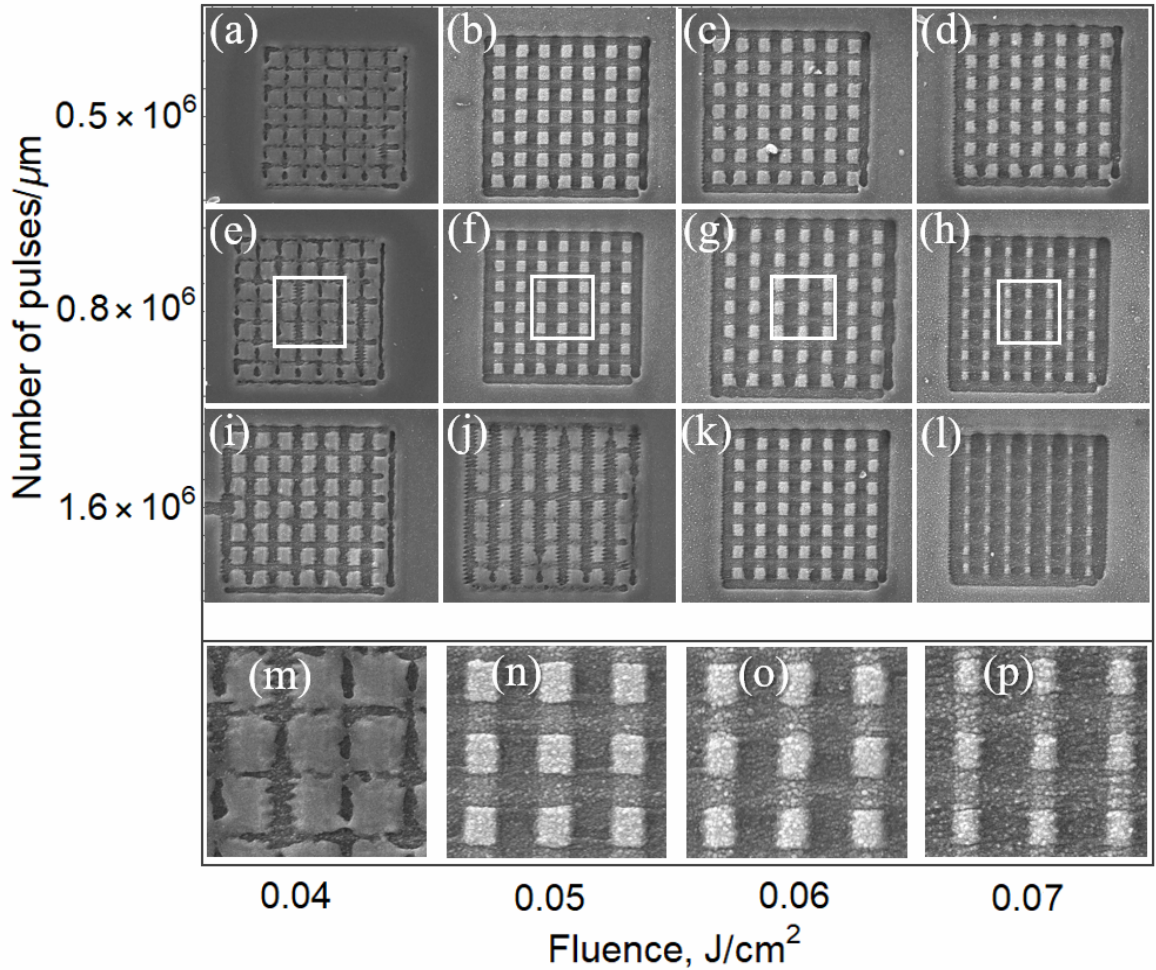

Fig. 5. (a - l) - SEM images of fishnet metasurfaces on a sapphire substrate with 3 μm period and different fabrication parameters. (m - p) – Magnified regions of the SEM images outlined in (e - h)

## Conclusion

In conclusion we introduce a powerful approach to fabrication of micro-structured tunable metasurfaces utilizing chalcogenide phase-change media and DLW technique. We have presented fishnet metasurfaces with period of 3 μm fabricated by laser scrabbing of GST film on glass and sapphire substrates. We found optimal modes for each substrate – for glass the fluence was $4.2 \times 10^2$ J/cm$^2$ and dose was $8 \times 10^5$ pulses per μm, when for sapphire substrate the fluence was $4.4 \times 10^2$ J/cm$^2$ and dose was the same. The time to manufacture one well-ordered 20x20 μm$^2$ metasurface with the number of square elements of the order of 50 was approximately 10 seconds. We found that the use of sapphire substrate provides better accuracy of metasurface segments than the glass one due to the high thermal conductivity of the material (25.2 W/(m*K) for sapphire versus 0.8 W/(m*K) for glass), and it guarantees immutability to slight variations of fabrication parameters. Further, the samples were experimentally investigated by SEM and AFM. Our approach provides a flexible platform for fast and precise manufacturing of all dielectric optically rewritable metasurfaces for applications in the field of telecommunications and information processing.

## Financial support

This work was supported by the Russian Science Foundation (Project 20-12-00272).

## Declaration of competing interest

The authors declare that they have no known competing financial interests or personal relationships that could have appeared to influence the work reported in this paper.

## Acknowledgments

The authors are grateful to P.A. Belov and A.B. Pevtsov for discussing the results.

## References

[1]  Zheludev, Nikolay I. "The road ahead for metamaterials." Science 328.5978 (2010): 582-583.

[2]  Liu, Yongmin, and Xiang Zhang. "Metamaterials: a new frontier of science and technology." Chemical Society Reviews 40.5 (2011): 2494-2507.

[3]  Makarov, Sergey V., et al. "Light‐Induced Tuning and Reconfiguration of Nanophotonic Structures." Laser & Photonics Reviews 11.5 (2017): 1700108.

[4]  Ou, Jun-Yu, et al. "An electromechanically reconfigurable plasmonic metamaterial operating in the near-infrared." Nature nanotechnology 8.4 (2013): 252-255.

[5]  Driscoll, Tom, et al. "Memory metamaterials." Science 325.5947 (2009): 1518-1521.

[6]  Kolobov, Alexander V., et al. "Understanding the phase-change mechanism of rewritable optical media." Nature materials 3.10 (2004): 703-708.

[7]  Shportko, Kostiantyn, et al. "Resonant bonding in crystalline phase-change materials." Nature materials 7.8 (2008): 653-658.

[8]  Olivieri, Anthony, et al. "Plasmonic nanostructured metal–oxide–semiconductor reflection modulators." Nano letters 15.4 (2015): 2304-2311.

[9]  Hsiao, Hui‐Hsin, Cheng Hung Chu, and Din Ping Tsai. "Fundamentals and applications of metasurfaces." Small Methods 1.4 (2017): 1600064.

[10]  Yang, Ki-Yeon, et al. "Patterning of Ge2Sb2Te5 phase change material using UV nano-imprint lithography." Microelectronic engineering 84.1 (2007): 21-24.

[11] Dong, Weiling, et al. "Tunable mid‐infrared phase‐change metasurface." Advanced Optical Materials 6.14 (2018): 1701346.

[12] Qian, Linyong, et al. "Tunable guided-mode resonant filter with wedged waveguide layer fabricated by masked ion beam etching." Optics Letters 41.5 (2016): 982-985.

[13] Rybin, Mikhail V., et al. "Band structure of photonic crystals fabricated by two-photon polymerization." Crystals 5.1 (2015): 61-73.

[14] Rybin, Mikhail V., et al. "Transition from two-dimensional photonic crystals to dielectric metasurfaces in the optical diffraction with a fine structure." Scientific reports 6 (2016): 30773.

[15] Yang, Liang, et al. "Parallel direct laser writing of micro-optical and photonic structures using spatial light modulator." Optics and Lasers in Engineering 70 (2015): 26-32.

[16] Chu, Cheng Hung, et al. "Active dielectric metasurface based on phase‐change medium." Laser & Photonics Reviews 10.6 (2016): 986-994.

[17] Karvounis, Artemios, et al. "All-dielectric phase-change reconfigurable metasurface." Applied Physics Letters 109.5 (2016): 051103.

[18] Gholipour, Behrad, et al. "Phase-change-driven dielectric-plasmonic transitions in chalcogenide metasurfaces." NPG Asia Materials 10.6 (2018): 533-539.

[19] Wei, Maoliang, et al. "Large-angle mid-infrared absorption switch enabled by polarization-independent GST metasurfaces." Materials Letters 236 (2019): 350-353.

[20] Bai, Wei, et al. "Near-infrared tunable metalens based on phase change material Ge 2 Se 2 Te 5." Scientific reports 9.1 (2019): 1-9.

[21] Wang, Qian, et al. "Optically reconfigurable metasurfaces and photonic devices based on phase change materials." Nature Photonics 10.1 (2016): 60-65.

[22] Li, Yaru, et al. "Switchable quarter-wave plate and half-wave plate based on phase-change metasurface." IEEE Photonics Journal 12.2 (2020): 1-10.

[23] Pogrebnyakov, Alexej V., et al. "Reconfigurable near-ir metasurface based on ge 2 sb 2 te 5 phase-change material." Optical Materials Express 8.8 (2018): 2264-2275.